\documentclass[conference]{IEEEtran}
\IEEEoverridecommandlockouts
\usepackage{cite}
\usepackage{amsmath,amssymb,amsfonts}
\usepackage{algorithmic}
\usepackage{graphicx}
\usepackage{textcomp}
\usepackage{xcolor}
\usepackage{tikz}
\usepackage{booktabs}
\usepackage{dblfloatfix}
\usepackage{placeins}
\usepackage{float}
\usetikzlibrary{arrows.meta, positioning, fit, backgrounds}
\makeatletter
\newcommand{\linebreakand}{%
  \end{@IEEEauthorhalign}%
  \hfill\mbox{}\par
  \mbox{}\hfill\begin{@IEEEauthorhalign}%
}
\makeatother

\def\BibTeX{{\rm B\kern-.05em{\sc i\kern-.025em b}\kern-.08em
    T\kern-.1667em\lower.7ex\hbox{E}\kern-.125emX}}

\begin{document}

\title{Adversarial RL for Port-Scan Evasion: Attacker Feature Visibility in Edge-Deployed IDS
}

\author{%
\IEEEauthorblockN{Logan Andrew North}
\IEEEauthorblockA{%
\textit{Pennsylvania State University}\\
University Park, PA, USA\\
lan5361@psu.edu}
\and
\IEEEauthorblockN{Priya Sanjay Kaluskar}
\IEEEauthorblockA{%
\textit{Pennsylvania State University}\\
University Park, PA, USA\\
psk5221@psu.edu}
\and
\IEEEauthorblockN{Shasi Kumar Ramachandran Prabhu}
\IEEEauthorblockA{%
\textit{Pennsylvania State University}\\
University Park, PA, USA\\
sfr5846@psu.edu}
%
\linebreakand
\IEEEauthorblockN{Peilong Li}
\IEEEauthorblockA{%
\textit{Elizabethtown College}\\
Elizabethtown, PA, USA\\
lip@etown.edu}
\and
\IEEEauthorblockN{Suman Saha}
\IEEEauthorblockA{%
\textit{Pennsylvania State University}\\
University Park, PA, USA\\
szs339@psu.edu}
}

\maketitle

\begin{abstract}
Machine learning-based intrusion detection systems (IDS) are increasingly used in resource-constrained Internet of Things (IoT) environments, yet their robustness is often evaluated against static attacks rather than adversaries that adapt to detection feedback. This paper investigates adaptive port-scan evasion against ML-based IDS models deployed on a Raspberry Pi 3B+. We implement a live Zeek-based IDS pipeline with XGBoost, a multi-layer perceptron, and a 1D convolutional neural network trained on TON\_IoT telemetry, and use a Deep Q-Network (DQN) adversary to learn evasive combinations of probe timing, TCP flags, and payload size under black-box, gray-box, and white-box feature-visibility settings. Although the deployed IDS models detect conventional port scans at 91.1--99.8\%, DQN final-50-episode evasion rates range from 61.9\% to 98.3\% across feature-visibility settings. Greater feature visibility does not monotonically improve evasion, and its effect is model-dependent: against XGBoost, the black-box agent achieves 92.9\% evasion, compared with 61.9\% and 76.9\% for gray-box and white-box agents, respectively, whereas 1D-CNN is most vulnerable under white-box access at 98.1\%. Because standard DQN can overestimate action values, we additionally
spot-check representative conditions using Double DQN. The gray-box
condition remains unstable in this check, providing no evidence that
overestimation bias alone explains the observed instability. These results show that limited feature knowledge can still enable effective adaptive evasion against static edge-deployed IDS models, motivating more robust defenses for IoT edge environments.
\end{abstract}

\section{Introduction}
IoT and edge deployments increasingly require intrusion detection on resource-constrained devices. Prior work has demonstrated the feasibility of deploying intrusion detection systems (IDS) on Raspberry Pi-class hardware~\cite{kyaw2015piids,sforzin2016rpids}, and machine learning (ML) is now widely used for detecting malicious network behavior beyond fixed signatures~\cite{buczak2015survey}. However, IDS robustness is still commonly evaluated against static attack traffic rather than adversaries that adapt in response to detection~\cite{apruzzese2022modeling,han2021evaluating}.

This gap matters especially for reconnaissance attacks such as port scanning, where an attacker can vary probing behavior over time. Prior work has shown ML-based IDSs are vulnerable to adversarial manipulation and has emphasized realistic black-box and gray-box threat models that preserve attack functionality~\cite{apruzzese2022modeling,han2021evaluating,jmila2022adversarial,debicha2023adv,merzouk2020deeper}. Reinforcement learning (RL) is a natural mechanism for such adaptation, since an agent can improve its policy through repeated interaction with a detector. Prior work has explored RL-based IDS evasion, including botnet-oriented attacks in dataset-driven settings~\cite{venturi2020drelab}, but the robustness of ML-based IDS deployed on real edge hardware against adaptive port-scan evasion -- particularly under varying attacker feature visibility -- remains insufficiently understood.

We investigate this using a live Raspberry Pi 3B+ IDS pipeline. Three models -- XGBoost~\cite{chen2016xgboost}, an MLP, and a 1D-CNN -- are trained on TON\_IoT telemetry~\cite{alsaedi2020toniot} and deployed with Zeek-based traffic monitoring~\cite{paxson1999bro}. A Deep Q-Network (DQN) adversary~\cite{mnih2015humanlevel}, implemented using Stable-Baselines3~\cite{raffin2021stable}, learns combinations of probe timing, TCP flags, and payload size under black-box, gray-box, and white-box feature-visibility settings.

This paper makes three contributions. First, it evaluates adaptive RL-based port-scan evasion against ML-based IDS running on real edge hardware, rather than only offline or simulated evaluation. Second, it examines how attacker feature visibility affects learned evasion across three IDS architectures. Third, it shows that strong conventional detection does not imply robustness to adaptive attack: final-50-episode evasion rates (the RL agent's success rate at getting a probe classified \texttt{normal}) range from 61.9\% to 98.3\% across evaluated conditions, and the effect of added feature visibility is model-dependent rather than monotonic -- XGBoost is most vulnerable black-box, while 1D-CNN is most vulnerable white-box. This exposes an attacker--defender asymmetry: even with limited feature visibility, the adversary improves through repeated interaction while the deployed detector stays static.

We scope this study as a vulnerability characterization, not a defense evaluation: we do not propose or test countermeasures here. Establishing that this evasion is practical against a live, edge-deployed IDS is a prerequisite to defending against it; candidate directions -- adaptive retraining, threshold randomization, and co-adaptive attacker--defender evaluation -- are discussed as future work in Section~V.

Section~II reviews related work; Section~III presents the system and methodology; Section~IV reports results; Section~V discusses limitations and future work; Section~VI concludes.

\section{Related Work}
\subsection{Intrusion Detection on Resource-Constrained Devices}
Low-cost edge hardware can support practical intrusion detection: Kyaw et al.~\cite{kyaw2015piids} evaluated Snort and Bro on a Raspberry Pi 2 against denial-of-service and port-scanning attacks, and Sforzin et al.~\cite{sforzin2016rpids} proposed RPiDS, a Raspberry Pi-based IDS architecture for IoT environments. ML is now widely applied to network intrusion detection~\cite{buczak2015survey}, supported by datasets such as TON\_IoT~\cite{alsaedi2020toniot} and Edge-IIoTset~\cite{ferrag2022edge}, and by work on compact feature representations for efficient IoT detection~\cite{li2024optimizing}. These studies establish the feasibility of lightweight, data-driven IDS deployment, but do not examine how an edge-deployed detector responds to an attacker that adapts its probing strategy through repeated feedback.

\subsection{Adversarial Attacks Against ML-Based Network IDS}
ML-based network IDS models are vulnerable to adversarial manipulation~\cite{jmila2022adversarial}. More recent work emphasizes realistic attacker assumptions: Apruzzese et al.~\cite{apruzzese2022modeling} highlight limitations of attacks assuming excessive model knowledge or infeasible feature manipulation, while Han et al.~\cite{han2021evaluating} evaluate practical black-box and gray-box attacks in traffic space; others examine functionality-preserving adversarial attacks and the practical validity of adversarial network traffic~\cite{debicha2023adv,merzouk2020deeper}.

Our work differs in deployment and attack adaptation: rather than evaluating only offline classifiers or fixed adversarial examples, we interact with ML-based IDS models running in a live edge-device pipeline and use RL to adapt port-scanning behavior from detector feedback. This assumes the adversary can repeatedly interact with the target during training -- an explicit threat-model choice most applicable to prolonged reconnaissance or repeated querying of a similar deployment, not a claim that applies to every deployment scenario.

\subsection{Reinforcement Learning for IDS Evasion and Defense}
The closest prior work is DReLAB~\cite{venturi2020drelab}, which uses deep RL (including DDQN and Deep Sarsa) to generate adversarial botnet samples against ML-based detectors, modeling the attacker as a learning agent rather than a fixed perturbation process, as we do. It focuses on botnet evasion and benchmark-data generation, however, whereas our work studies adaptive port-scan evasion through repeated interaction with IDS models deployed on Raspberry Pi edge hardware and explicitly compares attacker feature-visibility settings.

Prior work has also explored defenses such as hardening ML-based cyber detectors and adversarial training~\cite{apruzzese2020hardening,xiong2023aidtf} -- complementary to this study, which characterizes edge-deployed IDS vulnerability to an adaptive adversary rather than proposing a new defense.

\section{System and Methodology}

\begin{figure}[t]
    \centering
    \includegraphics[width=\columnwidth, trim=3pt 3pt 3pt 3pt, clip]{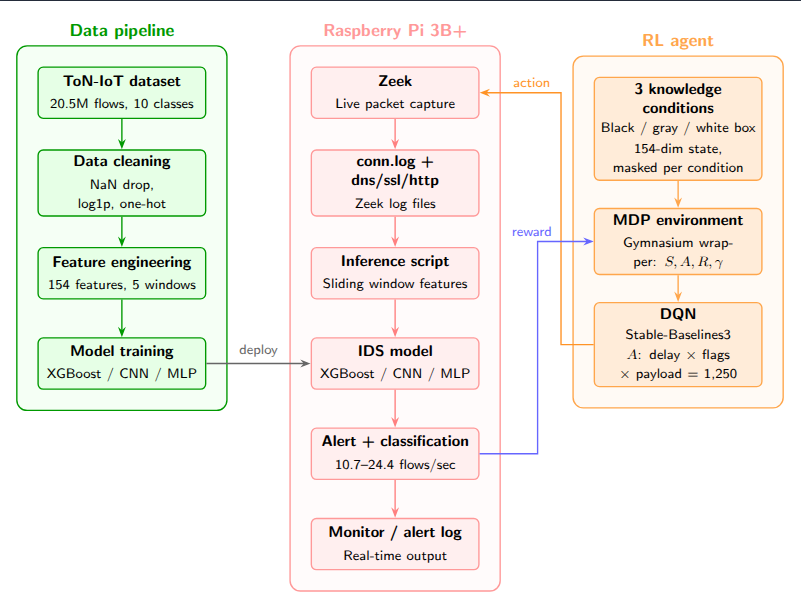}
    \caption{System architecture overview.}
    \label{fig:architecture}
\end{figure}

Our evaluation combines offline IDS training with live inference on a Raspberry Pi 3B+ and DQN-based adversarial probing. Figure~\ref{fig:architecture} summarizes the pipeline from feature construction and model training to live traffic monitoring, classification, and attacker feedback.

\subsection{System Overview}

Zeek monitors live traffic on the Raspberry Pi; the deployed IDS classifies each probe, and the prediction returns to the RL agent through a reward interface. The IDS stays fixed during an attack while the agent updates its policy over inter-probe delay, TCP flag, and payload size. The Pi's resource constraint applies only to the deployed IDS, not the learning process itself, so this setup tests whether a static, edge-deployed IDS remains robust to an attacker that learns from repeated interaction.

\subsection{Dataset and Feature Construction}

We train the IDS models on the TON\_IoT telemetry dataset~\cite{alsaedi2020toniot}, removing records with missing values and duplicates, applying \texttt{log1p} to numerical features, one-hot encoding categorical fields, and dropping high-cardinality or non-informative fields. To capture the temporal structure of port scanning, we augment connection-level features with sliding-window aggregates over 3/5/10/30/60\,s intervals (flow count, failed-connection ratio, port entropy, among others), yielding a 154-dimensional representation used consistently for offline training and live inference alike.

Models were trained on a 70/15/15 stratified train/validation/test split (seed 42). Sliding-window features were computed causally per source IP prior to splitting, so no feature uses information from a later connection, and we confirmed no direct label leakage. One caveat: windowing happens per source IP \emph{before} the random stratified split, so temporally adjacent rows from the same session -- with nearly identical, overlapping windows -- can land in different splits, a mild near-duplicate leakage distinct from label leakage. A grouped split would avoid this, but TON\_IoT's class-contiguous capture structure means it risks dropping entire attack classes from a partition, which we judged the worse risk. We therefore treat offline F1 as an upper bound rather than a leakage-free estimate, with the live Raspberry Pi results as the more trustworthy real-world indicator.

\subsection{Edge IDS Models}

We evaluate three IDS architectures: XGBoost, a 1D-CNN, and an MLP, all trained on the same 154-dimensional representation with class weighting and (for the neural models) early stopping. Table~\ref{tab:model-configs} summarizes architectures and training configurations. All three are deployed for live inference on the Raspberry Pi.

\begin{table}[t]
\centering
\caption{IDS model configurations.}
\label{tab:model-configs}
\begin{tabular}{p{0.18\linewidth} p{0.72\linewidth}}
\hline
\textbf{Model} & \textbf{Configuration} \\
\hline
XGBoost & 500 estimators, learning rate 0.05, maximum depth 6, subsampling ratio 0.8, L1 regularization with $\alpha=0.1$, class-weighted training, early stopping with patience 50. \\
1D-CNN & Two convolutional blocks with 32 and 64 filters, batch normalization and max pooling after each block, followed by a 128-unit dense layer, dropout 0.5, and softmax output. Trained with Adam on the full training set with class weights capped at $50\times$, learning-rate reduction on plateau, and early stopping with patience 15. \\
MLP & Four hidden layers with 64, 32, 16, and 8 neurons, ReLU activations, dropout 0.2 after the first layer, and softmax output. Trained with the same class-weighting and early-stopping strategy used for the 1D-CNN. \\
\hline
\end{tabular}
\end{table}

\subsection{Live IDS Deployment}

Our inference pipeline reconstructs the same 154-dimensional representation, including per-source sliding-window statistics, live from Zeek rather than relying on offline test-set classification alone, preserving the edge deployment's timing and feature-construction constraints. Predictions are filtered by source IP and matched to the agent's probe source port before reaching the reward socket; background traffic is classified but does not contribute to the agent's reward.

\subsection{RL-Based Port Scanning Agent}

We formulate port-scan evasion as a Markov Decision Process (MDP), $\langle S,A,T,R,\gamma\rangle$~\cite{sutton2018reinforcement}, sequential because each probe changes the sliding-window statistics that influence subsequent IDS observations:

\begin{itemize}
    \item $S$: a 154-dimensional observation aligned with the IDS feature representation -- all-zero, partially masked, or complete, depending on the attacker feature-visibility setting.

    \item $A$: $50 \times 5 \times 5 = 1{,}250$ discrete combinations of inter-probe delay (50 log-spaced values, 0.1--5.0\,s), TCP flag (\{SYN, FIN, NULL, Xmas, SYN-ACK\}), and payload size (\{0, 64, 128, 256, 512\} bytes).

    \item $T$: transition dynamics induced by the live network and IDS pipeline, where each probe and response update the sliding-window features used in subsequent steps.

    \item $R$: $+1$ minus a delay penalty when classified normal, $-1$ when classified scanning, $0$ otherwise.

    \item $\gamma$: discount factor $0.99$, over each 20-probe episode.
\end{itemize}

\[
R(s,a)=
\begin{cases}
+1 - \frac{d}{d_{\max}}\cdot \omega, & \text{if classified as normal},\\
-1, & \text{if classified as scanning},\\
0, & \text{otherwise},
\end{cases}
\]
where $d_{\max}=5.0$\,s and $\omega$ controls the penalty for slower probing (Section~\ref{sec:sensitivity}).

\subsection{Reward Sensitivity Analysis}
\label{sec:sensitivity}

We evaluate $\omega \in \{0.5,1.0,2.0\}$ using the XGBoost black-box condition (300 episodes / 6{,}000 timesteps each). At $\omega=1.0$, selecting the maximum delay yields zero reward even when classified normal -- the theoretical breakeven point against maximum-delay stalling. Empirically, $\omega=1.0$ and $\omega=2.0$ converge faster and more stably than $\omega=0.5$ with comparable final evasion, so we use $\omega=1.0$ throughout; Table~\ref{tab:hyperparams} summarizes the remaining DQN hyperparameters.

\begin{table}[t]
\centering
\caption{DQN Hyperparameters.}
\label{tab:hyperparams}
\begin{tabular}{lc}
\toprule
\textbf{Hyperparameter} & \textbf{Value} \\
\midrule
Discount factor $\gamma$         & 0.99 \\
Learning rate $\alpha$           & $10^{-4}$ \\
Replay buffer size               & 10{,}000 \\
Batch size                       & 32 \\
Online network update frequency  & every 4 steps \\
Target network update frequency  & every 100 steps \\
Exploration $\varepsilon$ (start) & 1.0 \\
Exploration $\varepsilon$ (end)   & 0.05 \\
Exploration decay                & first 30\% of timesteps \\
Delay penalty weight $\omega$    & 1.0 \\
Training timesteps               & 10{,}000 \\
Episode length                   & 20 probes \\
\bottomrule
\end{tabular}
\end{table}

Main experiments use 10{,}000 timesteps (500 episodes) per target model and feature-visibility setting; the 6{,}000-timestep runs above were used only to select $\omega$.

\subsection{Attacker Feature Visibility}
\label{sec:gray-box-features}

We evaluate three attacker feature-visibility settings, varying access to the IDS feature representation rather than to the target model's parameters or architecture.

\begin{itemize}

\item \textbf{Black-box.} The agent's observation is an all-zero vector at every timestep, so the policy is stateless -- though the environment still evolves, since the IDS maintains sliding-window statistics. We treat this as bandit-like from the agent's perspective while retaining DQN for consistency across settings.

\item \textbf{Gray-box.} The agent observes only the top $k=20$ features for the target model, all other dimensions zeroed. Top-20 sets were derived independently per model: for XGBoost, by gain importance computed directly from the trained model (training-data-derived, since gain reflects the splits made while fitting on \texttt{X\_train}); for 1D-CNN and MLP, which have no native split-based importance measure, by mean absolute SHAP value~\cite{lundberg2017unified} (\texttt{shap.DeepExplainer}, 100 background / 500 evaluation samples from the test set).

\item \textbf{White-box.} The agent receives the complete 154-dimensional feature vector -- full feature visibility only, not access to model parameters or training procedure.

\end{itemize}

All target-model and visibility combinations use the same DQN architecture and the main training budget of 10{,}000 timesteps (500 episodes).

\section{Experimental Results}

We first establish each deployed IDS model's ability to detect conventional port scans and then evaluate its robustness to adaptive RL-based evasion.

\subsection{Baseline IDS Performance}

Before adversarial training, each model was evaluated against a standard \texttt{nmap -sS} SYN scan~\cite{lyon2009nmap} in the live Raspberry Pi deployment. Table~\ref{tab:baseline} reports offline macro F1 and scanning-class F1, live detection rate, inference throughput, and model size. All three models detect conventional scan traffic effectively (live detection rates 91.1--99.8\%; throughput 10.7--24.4 flows/s; model sizes 52.8~KB--14.5~MB), establishing that subsequent evasion is measured against models that can both detect standard scans and operate within edge-device constraints.

\begin{table}[t]
\centering
\caption{Baseline IDS model performance against standard \texttt{nmap -sS} port scan (1{,}000 ports). Scanning F1 and Macro F1 computed from offline test set; Live Detection and Flows/sec measured during live deployment on the Raspberry Pi 3B+. F1 scores rounded to 3 decimal places.}
\label{tab:baseline}
\resizebox{\columnwidth}{!}{%
\begin{tabular}{lccccc}
\toprule
\textbf{Model} & \textbf{Macro F1} & \textbf{Scanning F1} & \textbf{Live Detection} & \textbf{Flows/sec} & \textbf{Size} \\
\midrule
XGBoost & 0.997 & 1.000 & 99.8\% & 10.7 & 14.5 MB \\
1D-CNN  & 0.981 & 1.000 & 98.0\% & 23.1 & 1.22 MB \\
MLP     & 0.923 & 0.990 & 91.1\% & 24.4 & 52.8 KB \\
\bottomrule
\end{tabular}%
}
\end{table}

\subsection{Adversarial Evasion Results}

Table~\ref{tab:evasion-summary} summarizes DQN evasion performance across target models and attacker feature-visibility settings at $\omega=1.0$ over 500 episodes, reported as mean $\pm$ standard deviation across seeds 42, 67, and 320. These figures reflect how each classifier scored the actions it was given, independent of the shared-seed effect discussed in Section~\ref{sec:strategies}.

The effect of feature visibility is strongly model-dependent. XGBoost is most vulnerable under black-box access (92.9\% $\pm$ 2.8\%), dropping to a lower and far more variable 61.9\% $\pm$ 19.0\% under gray-box. 1D-CNN, in contrast, is most vulnerable under white-box visibility (98.1\% $\pm$ 0.4\%). MLP remains highly vulnerable across all three settings (96.3--98.3\%). Greater feature visibility does not monotonically improve adaptive evasion.

\begin{figure*}[t]
    \centering
    \includegraphics[width=\textwidth]
    {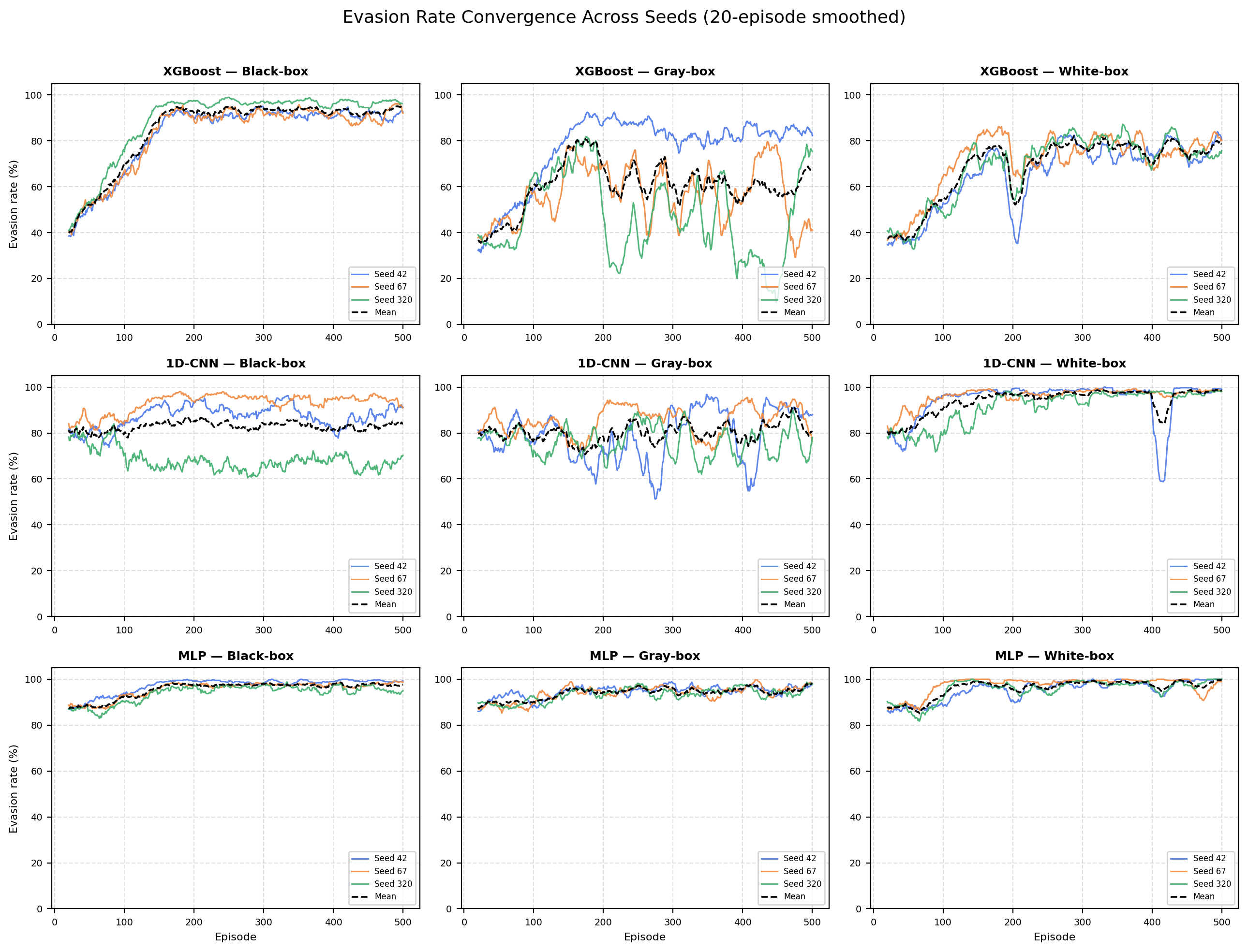}
    \caption{Evasion-rate convergence across target models, attacker
    feature-visibility settings, and available seeds (20-episode moving
    average). The dashed black line shows the mean across available
    seeds when multiple runs are present.}
    \label{fig:convergence}
\end{figure*}

Figure~\ref{fig:convergence} shows convergence trajectories across available seeds; XGBoost gray-box exhibits the largest cross-seed instability (Section~\ref{sec:convergence}).

\begin{table}[t]
\centering
\caption{DQN evasion performance and representative dominant strategy
across IDS models and attacker feature-visibility settings
($\omega=1.0$, 500 episodes). Avg.\ Last 50 is reported as mean
$\pm$ standard deviation across available seeds. Dominant strategies
are from seed 42 and are reported as flag / delay / payload;
``variable'' indicates that no single value exceeded 40\% of
post-exploration actions.}
\label{tab:evasion-summary}
\resizebox{\columnwidth}{!}{%
\begin{tabular}{llcl}
\hline
\textbf{Target} & \textbf{Visibility} &
\textbf{Avg. Last 50} & \textbf{Dominant Strategy} \\
\hline
XGBoost & Black-box  & $92.9 \pm 2.8\%$  & SYN / 0.26s / 512B \\
XGBoost & Gray-box   & $61.9 \pm 19.0\%$ & SYN-ACK / 0.11s / 512B \\
XGBoost & White-box  & $76.9 \pm 2.3\%$  & SYN-ACK / variable / 512B \\
1D-CNN  & Black-box  & $84.3 \pm 12.3\%$ & SYN / 0.58s / 64B \\
1D-CNN  & Gray-box   & $84.9 \pm 3.9\%$  &
variable / variable / 512B \\
1D-CNN  & White-box  & $98.1 \pm 0.4\%$  &
SYN-ACK / variable / 512B \\
MLP     & Black-box  & $97.6 \pm 1.8\%$  & SYN / 0.58s / 64B \\
MLP     & Gray-box   & $96.3 \pm 0.5\%$  &
SYN-ACK / variable / 512B \\
MLP     & White-box  & $98.3 \pm 1.8\%$  & SYN / variable / 512B \\
\hline
\end{tabular}%
}
\end{table}

\subsection{Learned Evasion Strategies}
\label{sec:strategies}

Under black-box visibility, all three models favor SYN probes with sub-second delays and fixed payloads (96.1\%, 95.9\%, 96.1\% concentration for XGBoost, 1D-CNN, MLP). This close concentration revealed a broader finding about DQN under shared random seeds: our training harness seeds network initialization, exploration, replay sampling, and port-queue shuffling identically across target models for a given seed. A genuinely different reward signal should still drive each model to its own solution -- but under black-box visibility, where every model's observation is a constant all-zero vector, it does not: at seeds 67 and 320, XGBoost, 1D-CNN, and MLP take \emph{identical} action sequences step-for-step for the entire run, despite each classifier scoring those actions differently (reward agreement between XGBoost and 1D-CNN is only 78\%).

This is specific to the DQN configuration evaluated here, not the reward design: an $\epsilon$-greedy multi-armed bandit trained under the identical shared-seed setup (Section~\ref{sec:ddqn}) diverges cleanly across models instead (8.5\% action agreement, comparable to gray-box's independent 10--12\% baseline). We cannot isolate which specific difference is responsible, since DQN and the bandit differ along several axes at once -- batched gradient updates (batch size 32, learning rate $10^{-4}$, every 4 steps, $\sim$2,500 updates total) versus the bandit's direct, single-step update $Q[a] \leftarrow Q[a] + (r-Q[a])/n[a]$; a bootstrapped target, $r+\gamma\max_{a'}Q(s',a')$, versus the bandit's immediately-observed reward; neural function approximation versus per-arm tabular values -- and we ran no ablation varying these individually. Batching and bootstrapping are plausible contributing factors -- both dilute or add noise to a 78\%-agreeing reward signal in ways the bandit's update does not -- but we treat them as candidate explanations, not a confirmed cause. What the comparison does establish: under this shared-seed setup, DQN does not diverge across target models in the black-box condition, while the bandit does. We report black-box dominant strategies in Table~\ref{tab:evasion-summary} as a single, largely shared action sequence evaluated against three classifiers, not three independent discoveries.

With richer visibility the strategies become model-dependent: XGBoost shifts to SYN-ACK under gray-box and white-box (73\%, 61\%); 1D-CNN favors SYN-ACK under white-box (72\%) but is more diffuse under gray-box; MLP favors SYN-ACK in gray-box and reverts to SYN in white-box. Gray-box escapes the shared-seed effect above -- each model has a structurally distinct top-20 feature mask (Section~\ref{sec:gray-box-features}) from the first step, and post-exploration actions diverge sharply across models (10--12\% agreement) onto genuinely different dominant flags. White-box, giving every model the identical full vector, is exposed to the same effect as black-box, though less severely at seed 42 (23.3\%/7.7\% agreement) than at seed 67 (47--48\%, all three converging on SYN-ACK). We treat gray-box as the strongest evidence here for genuinely independent, model-specific strategy learning.

\textbf{Reconnaissance validity:} IDS evasion does not necessarily imply useful reconnaissance, and leaving the action space unconstrained let the agent reveal this distinction rather than assume it away. This evidence rests on the environment's own logged \texttt{port\_state} field, not independently-audited ground truth for every probe; we found the field unreliable in one case (below), so we treat these conclusions as evidence rather than certainty. We draw on every probe log recording raw TCP responses -- five cross-model transfer runs (Section~\ref{sec:transfer}) and three Double DQN spot-check runs (Section~\ref{sec:doubledqn-check}), 32,500 probes total. The bandit baseline (Section~\ref{sec:ddqn}) does not log this field, so its NULL-dominant black-box strategy cannot be verified the same way; the flag-level findings below inform its interpretation, not confirm it.

SYN probes align with conventional half-open scanning~\cite{lyon2009nmap}: 19,394 of 19,436 closed-port hits return \texttt{RST-ACK}, and 42 land on genuinely open ports, returning the expected \texttt{SYN-ACK} -- differential, reconnaissance-retaining behavior at scale. SYN-ACK probes, by contrast, elicit a uniform \texttt{RST} on all 5,940 instances regardless of logged port state -- no evidence of retained reconnaissance utility, whatever its evasion rate. FIN (n=2,320) and NULL (n=2,654) show the same differential pattern as SYN: \texttt{RST-ACK} on closed ports (2,313 and 2,645), with a small number of \texttt{no\_response} instances (7 and 9) consistent with RFC~793's silent-drop behavior on non-closed ports. NULL was previously unverified here; this larger pool now shows it retains reconnaissance value like SYN and FIN. Xmas (n=2,150) shows the identical pattern, including one instance independently confirmed against the live host (port 111, listening at scan time, environment-logged as \texttt{filtered} rather than \texttt{open} -- a logging limitation, not a stack deviation, and the only instance checked this way). Given NULL's confirmed value, the bandit's NULL-dominant black-box strategy is plausibly reconnaissance-retaining too, though not independently verified for its own probes, since its logs lack the response fields; we infer this only from NULL's behavior elsewhere in this pool.

This is most visible for XGBoost: the shift from SYN toward SYN-ACK under richer visibility costs the agent on both axes at once -- lower evasion \emph{and} no observed port-state discrimination -- showing that more information does not automatically produce a better decision once "better by the reward function" and "better for the attacker's goal" can diverge. This RST pattern is a fixed property of the flag against this stack, regardless of the single-run caveat above; the relationship does not hold for 1D-CNN or MLP, so whether richer visibility produces this trade-off is itself model-dependent.

\begin{figure*}[t]
    \centering
    \includegraphics[width=\textwidth]{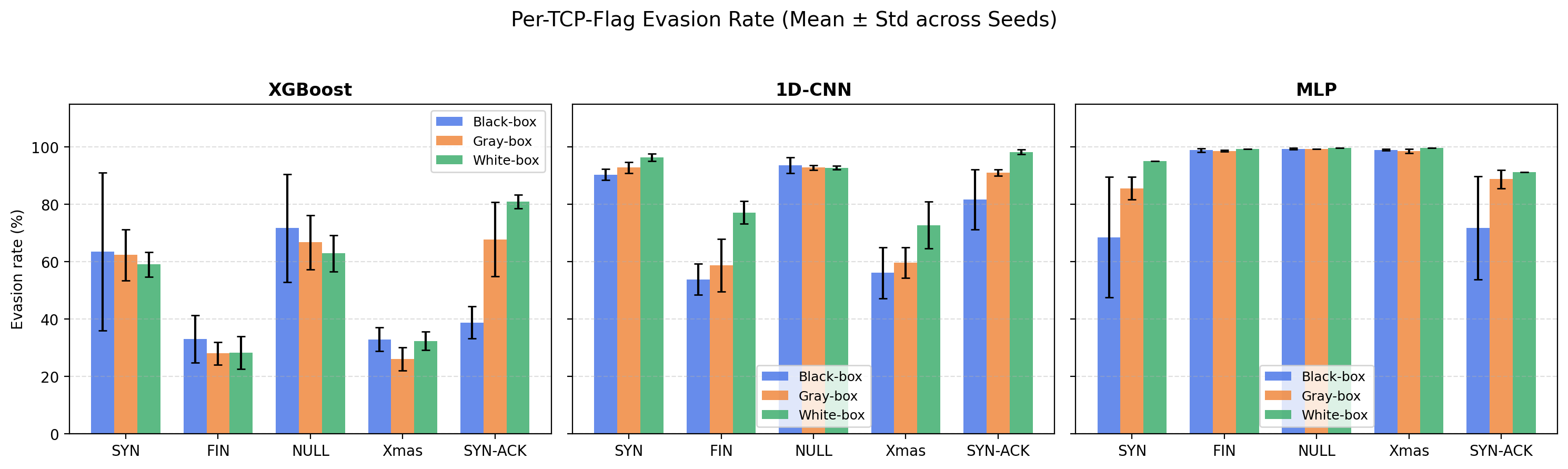}
    \caption{Per-TCP-flag evasion rate by target model and attacker
    feature-visibility setting (mean $\pm$ standard deviation across
    available seeds). Error bars indicate cross-seed variability; XGBoost
    gray-box exhibits the largest variance, consistent with the convergence
    behavior shown in Figure~\ref{fig:convergence}.}
    \label{fig:flag_evasion}
\end{figure*}

Figure~\ref{fig:flag_evasion} shows per-flag evasion across seeds: for XGBoost, SYN-ACK evasion rises substantially under gray/white-box -- the clearest case of a strategy excellent by reward with no observed reconnaissance value. 1D-CNN and MLP stay high across several flags, indicating less dependence on any one.

\subsection{DQN versus a Multi-Armed Bandit Baseline}
\label{sec:ddqn}

Black-box's all-zero observation means DQN cannot condition on IDS feature state; we treat it as bandit-like from the agent's perspective while retaining DQN throughout for architectural consistency.

To test whether a simpler stateless learner can exploit this, and to probe the shared-seed effect directly, we train an $\epsilon$-greedy multi-armed bandit over the same 1,250-action space and reward, without state or replay, for 500 episodes per target model at seed 42. Each action is an independent arm with value $Q[a]$ updated by incremental mean, $Q[a]\leftarrow Q[a]+(r-Q[a])/n[a]$; the highest-value arm is chosen with probability $1-\epsilon$, a uniform-random arm otherwise, $\epsilon$ decaying linearly from 1.0 to 0.05 over the first 30\% of timesteps to mirror Stable-Baselines3 DQN's defaults. The bandit ignores the observation entirely, and its three per-model runs are independent under the shared seed (8.5\% action agreement), unlike DQN's.

The bandit reaches final-50-episode evasion of 97.4\% (XGBoost), 99.8\% (1D-CNN), and 98.1\% (MLP) -- single-seed results not directly comparable, statistically, to the multi-seed DQN figures in Table~\ref{tab:evasion-summary}. The two learners also favor different strategies: DQN favors SYN across all three black-box targets, while the bandit favors NULL (XGBoost, MLP) and SYN-ACK (1D-CNN); we do not treat this as evidence either learner's strategy is more reconnaissance-useful, since flag-level reconnaissance utility was not checked for every flag (Section~\ref{sec:strategies}). This is one largely-shared DQN result set against three independently-converged bandit results, not a symmetric comparison -- showing the reduced cross-model divergence is specific to the evaluated DQN configuration, not the task or reward, though we do not claim to have isolated which particular difference (update rule, bootstrapping, function approximation, or some combination) is responsible.

\begin{figure*}[t]
    \centering
    \includegraphics[width=\textwidth]{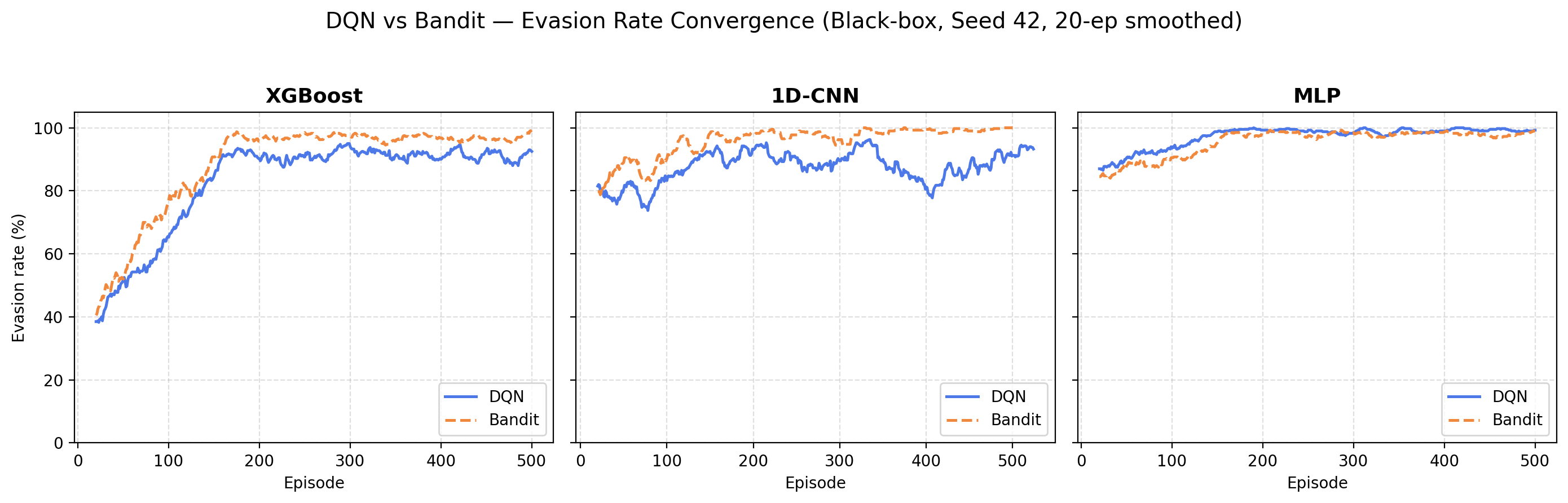}
    \caption{DQN versus multi-armed bandit evasion-rate convergence under
    black-box feature visibility for seed 42 (20-episode moving average).
    The comparison is descriptive because the bandit ablation is limited
    to one seed.}
    \label{fig:bandit}
\end{figure*}

Figure~\ref{fig:bandit} illustrates the seed-42 convergence behavior. The ablation shows DQN is not necessary for high evasion in the zero-observation setting, and is consistent with a difference in how the two learners respond to reward under shared initial conditions; we retain DQN throughout for a common learning framework across all three feature-visibility settings.

\subsection{Convergence Stability and Seed Sensitivity}
\label{sec:convergence}

XGBoost gray-box shows the largest cross-seed variability (std 19.0\%, seed range 37.8--84.2\%); 1D-CNN black-box is next (12.3\%); all remaining conditions are at or below 3.9\%. Figure~\ref{fig:convergence} shows why: XGBoost gray-box seed 42 converges above 80\% by episode 100 and stays stable, while seeds 67 and 320 undergo extended degradation before partially recovering. Policy learning under partial feature visibility is particularly seed-sensitive for XGBoost; the present experiments do not isolate the mechanism, so we do not attribute it to a specific property of the masked representation.

\subsection{Double DQN Spot Check: Is Overestimation Bias a Factor?}
\label{sec:doubledqn-check}

Standard DQN is known to overestimate action values, an effect that can be more pronounced with larger action spaces such as ours; Double DQN~\cite{hasselt2016double} corrects for this by decoupling action selection (online network) from evaluation (target network). We retrained three representative conditions with Double DQN at seed 42, matching the 10,000-timestep budget of the corresponding standard-DQN runs (Table~\ref{tab:evasion-summary}), to check whether this concern affects our conclusions.

\begin{figure*}[t]
    \centering
    \includegraphics[width=\textwidth]{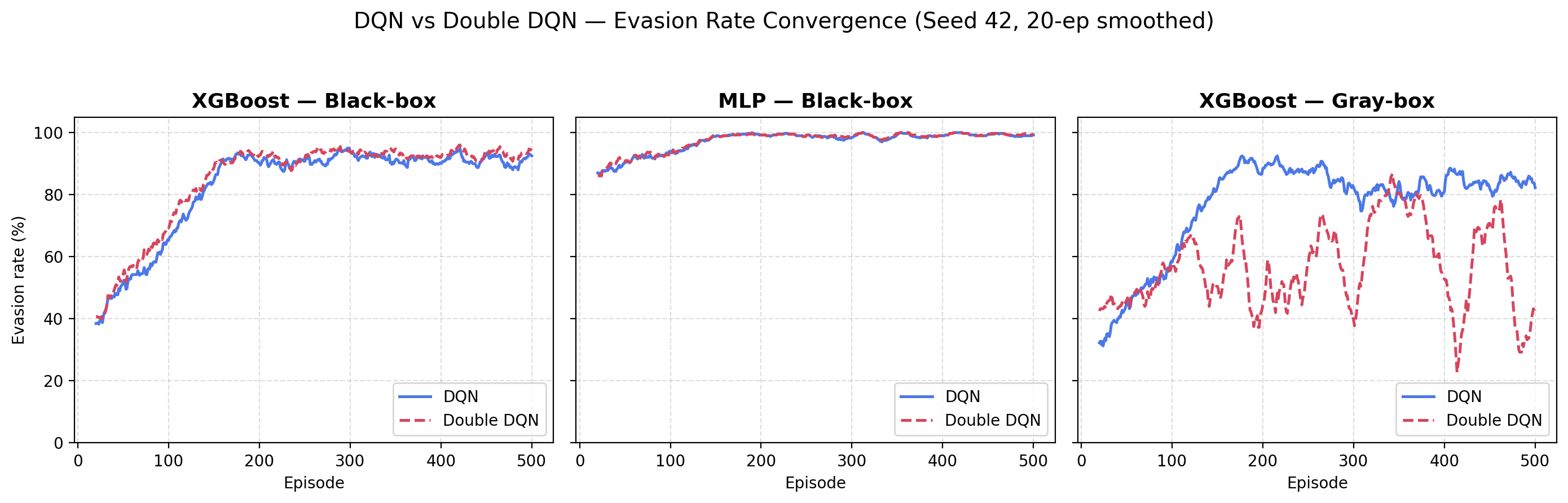}
    \caption{Standard DQN versus Double DQN evasion-rate convergence for three representative conditions (seed 42, 20-episode smoothed). Black-box trajectories closely track each other for both target models; gray-box remains unstable under both algorithms, with Double DQN showing more pronounced oscillation at the one seed where standard DQN was most stable.}
    \label{fig:doubledqn}
\end{figure*}

Black-box is nearly unaffected: XGBoost reaches 93.0\% avg.\ last-50 evasion under Double DQN versus 90.7\% under standard DQN, and MLP reaches 99.4\% versus 99.2\%, with closely overlapping trajectories (Figure~\ref{fig:doubledqn}, left/center) -- consistent with the constant all-zero observation giving overestimation little to distort. The dominant strategies sharpen the shared-seed finding from Section~\ref{sec:strategies}: both models converge to identical flag usage (SYN at 96.1\%, matching secondary-flag percentages to the decimal), but their delay/payload combinations differ and exactly match each model's existing standard-DQN values (XGBoost 0.26s/512B, MLP 0.58s/64B) -- the shared seed dominates flag selection almost completely while leaving room for genuine per-model differentiation in delay and payload.

Gray-box instability is not resolved -- if anything it worsens. Seed 42, standard DQN's \emph{most stable} gray-box seed (84.2\% avg.\ last-50, versus 37.8\% and 63.7\% for seeds 67/320), drops to 46.2\% under Double DQN and oscillates throughout training rather than plateauing (Figure~\ref{fig:doubledqn}, right). The strategy itself grows less decisive too: standard DQN's SYN-ACK-concentrated policy (73.2\% flag, 46.0\% delay) becomes far more diffuse under Double DQN (SYN-ACK 36.5\%, NULL 20.3\%, SYN 17.1\%, FIN 14.4\%, Xmas 11.8\%; no delay value above 10.2\%), with only payload (512B, 42.3\%) staying concentrated. If overestimation bias drove gray-box's seed sensitivity (Section~\ref{sec:convergence}), correcting it should have calmed this already-favorable seed; instead both evasion rate and strategy concentration got worse. This is more consistent with one candidate explanation already on the table: the partially-masked observation shifting as the agent's own evolving policy alters the sliding-window statistics it is conditioned on, rather than overestimation bias specifically -- Double DQN removes the latter but not the former.

Each condition uses a single Double DQN seed, and standard DQN's own gray-box results already span 37.8--84.2\% across seeds, so we cannot rule out this seed simply drawing an unfavorable trajectory. This is not a controlled ablation isolating overestimation bias, only evidence against it being the dominant explanation for Section~\ref{sec:convergence}'s instability; a multi-seed Double DQN comparison is left to future work.

\subsection{Time-to-First-Evasion}
\label{sec:tte}

\begin{figure*}[t]
    \centering
    \includegraphics[width=\textwidth]{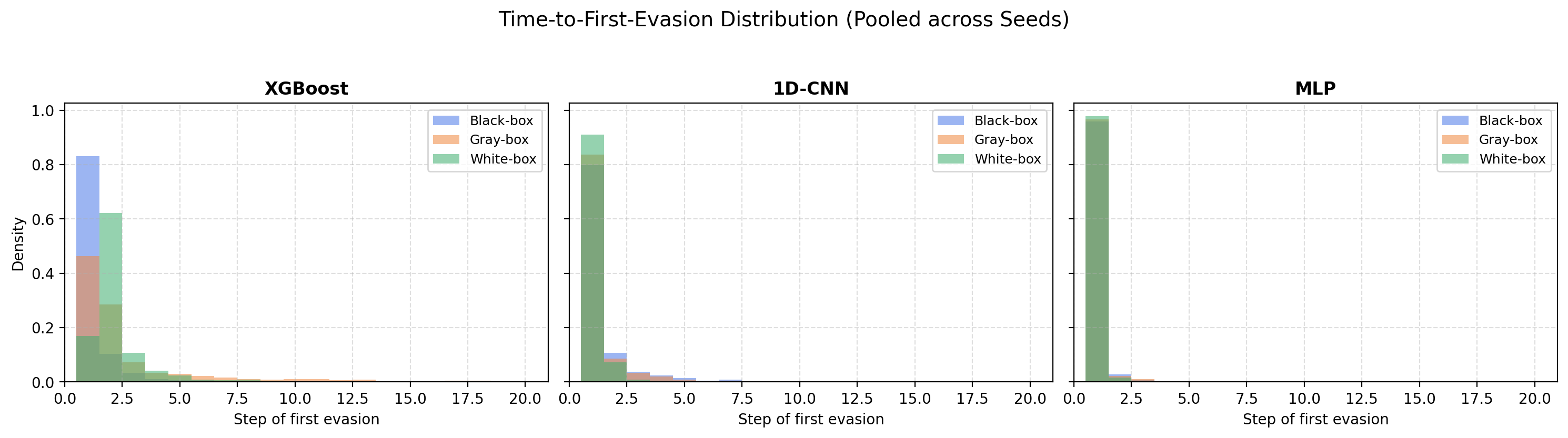}
    \caption{Time-to-first-evasion distribution across available seeds, measured as the step within each episode of the first successful evasion. Successful evasion is concentrated near the earliest episode steps across models and attacker feature-visibility settings.}
    \label{fig:tte}
\end{figure*}

Figure~\ref{fig:tte} shows the step of first evasion, pooled across seeds. 1D-CNN and MLP evade early and consistently: MLP reaches step-1 evasion in 95.8--97.7\% of episodes across all three settings, and 1D-CNN in 79.8--90.9\%. XGBoost is markedly more spread: black-box reaches step-1 evasion in 83.2\% of episodes (mean step 1.32), but gray-box and white-box are both far more distributed (mean steps 2.72 and 2.33, with only 46.4\% and 16.9\% of episodes evading at step 1, respectively) -- consistent with XGBoost gray-box's broader instability documented in Section~\ref{sec:convergence}.

\subsection{Cross-Model Transferability}
\label{sec:transfer}

The strategies above were each learned against a single target model. To test generalization to a different classifier, we freeze five trained seed-42 policies and evaluate each deterministically for 25 episodes against a different target under the same feature-visibility setting: XGBoost's black-, gray-, and white-box policies against 1D-CNN, 1D-CNN's white-box policy against XGBoost, and MLP's gray-box policy against XGBoost. For gray-box transfer, the frozen policy is evaluated using the destination classifier's own top-20 feature mask (Section~\ref{sec:gray-box-features}), not the mask it trained under -- since XGBoost's and 1D-CNN's masks share only 6 of 20 indices, the policy's input can differ substantially from what it saw during training.

Transfer performance is strongly source-target dependent. XGBoost's black-, gray-, and white-box policies achieve 99.8\%, 99.8\%, and 96.8\% evasion against 1D-CNN -- the black-box figure carries the same shared-seed caveat as its dominant-strategy result (Section~\ref{sec:strategies}), since it reflects a largely non-independent action sequence rather than an XGBoost-specific policy. The 1D-CNN white-box policy achieves 65.0\% against XGBoost, and the MLP gray-box policy achieves only 0.4\%. Policies learned against one IDS architecture do not transfer uniformly to another; the present experiments do not isolate the mechanism responsible.

\subsection{Discussion: Model-Dependent Effects of Attacker Feature Visibility}
\label{sec:double-edged}

The effect of attacker feature visibility (Table~\ref{tab:evasion-summary}) is non-monotonic and model-dependent: XGBoost is most vulnerable black-box and least under gray-box, 1D-CNN is most vulnerable white-box, and MLP stays near ceiling throughout. More IDS feature information does not necessarily produce stronger evasion.

The learned policies add context: in the seed-42 run, XGBoost shifts from SYN under black-box to SYN-ACK under gray/white-box. Since the reward never explicitly favors particular flags, this shift emerges from interaction with the detector rather than a predefined preference -- though, per Section~\ref{sec:strategies}, it should be read as a single-run illustration rather than a seed-replicated effect, with the evasion-rate contrast above as the load-bearing evidence.

From a security perspective, this cautions against assuming that limiting an attacker's feature access prevents adaptive evasion: even a black-box attacker learns effective probing through repeated feedback, while a white-box attacker's extra information does not guarantee a stronger attack, since its value depends on the target model and the policy learned. Robustness evaluation should consider adaptive attacker--detector interaction alongside conventional detection accuracy and attacker-knowledge assumptions.

\section{Limitations and Future Work}
Several limitations apply. The experiments consider three ML-based IDS models deployed on a Raspberry Pi 3B+ using a TON\_IoT-derived feature representation, so findings do not necessarily generalize to other edge platforms, datasets, or IDS architectures. The adversarial objective targets the \texttt{scanning} class specifically and should not be assumed to transfer to other attack types. Response-pattern evidence for reconnaissance validity (Section~\ref{sec:strategies}) covers all five learned probe flags, but rests mainly on the environment's own logged fields rather than independently-audited ground truth for every probe -- only one instance was cross-checked directly against the live host, and the bandit's NULL-dominant black-box strategy was not verified this way, since its logs lack the necessary fields.

Future work should evaluate adaptive defenses for edge-deployed IDS -- threshold randomization, adaptive/periodic retraining, and online adaptation -- alongside hardening and adversarial-training approaches~\cite{apruzzese2020hardening,xiong2023aidtf}. Since the adversary adapts here while the IDS stays fixed, co-adaptive attacker--defender experiments would show whether these evasion strategies persist against an evolving detector.

The transfer experiments in Section~\ref{sec:transfer} are limited to a small set of source-target pairs and seed-42 policies; extending this across more model pairs, seeds, feature-visibility settings, attack classes, datasets, and edge platforms would clarify which vulnerabilities are model-specific and which generalize.

\section{Conclusion}
This paper evaluated DQN-based adaptive port-scan evasion against ML-based IDS deployed on Raspberry Pi edge hardware. Despite strong conventional scan detection, final-50-episode evasion rates ranged from 61.9\% to 98.3\% across models and attacker feature-visibility settings, and the effect of added visibility was non-monotonic and model-dependent -- XGBoost most vulnerable black-box, 1D-CNN white-box, MLP highly evadable throughout. Strong baseline detection does not imply robustness to an adaptive adversary: limited feature visibility can still support effective evasion through repeated interaction with a static detector, motivating development of adaptive, adversarially robust IDS mechanisms for resource-constrained IoT and edge environments.

\bibliographystyle{IEEEtran}
\bibliography{main}

\end{document}